\documentclass[final,3p,times,twocolumn]{elsarticle}

\usepackage{amssymb}
\usepackage{amsmath}
\usepackage[
    colorlinks=true,
    linkcolor=blue,
    citecolor=blue,
    urlcolor=blue
]{hyperref}

\journal{Materials Today Quantum}

\begin{document}

\begin{frontmatter}

\title{Sidewall spacer passivated epitaxial NbN/AlN/NbN Josephson junctions for superconducting qubit applications}

\author[Tohoku]{Koki Honda} 
\ead{koki.honda.s6@dc.tohoku.ac.jp}
\author[Tohoku]{Daiki Kurihara}
\author[Tohoku]{Hiroki Kutsuma}
\author[NICT]{Hirotaka Terai}
\author[Tohoku]{Taro Yamashita}

\affiliation[Tohoku]{
    organization={Department of Applied Physics, Graduate School of Engineering, Tohoku University},
    city={Sendai},
    postcode={980-8579},
    state={Miyagi},
    country={Japan}}

\affiliation[NICT]{
    organization={Advanced ICT Research Institute, National Institute of Information and Communications Technology},
    city={Kobe},
    postcode={651-2492},
    state={Hyogo},
    country={Japan}}

\begin{abstract}
Epitaxial NbN/AlN/NbN Josephson junctions are promising alternatives to conventional Al/$\rm AlO_x$/Al junctions for superconducting qubits due to their crystalline tunnel barriers and the high transition temperature of NbN. However, their integration into quantum circuits typically requires complex fabrication processes involving amorphous interlayer dielectrics, which can degrade qubit coherence times. To address these issues, we developed sidewall spacer passivated epitaxial NbN/AlN/NbN junctions using a fabrication process that reduces both the volume of amorphous dielectric material and fabrication complexity. The fabricated junctions exhibited a large gap voltage of approximately 5.6 mV and low subgap leakage currents. By varying the thickness of the AlN tunnel barrier, high-quality junctions were fabricated over a wide range of critical current densities $J_{\rm c}$, from 0.3 to $10^4~\rm A/cm^2$. In addition, the comparison of the current–voltage characteristics before and after buffered hydrofluoric acid treatment indicated that the wet etching process did not degrade the electrical characteristics of the junctions. These results establish a scalable fabrication process for epitaxial Josephson junctions and provide a promising platform for high-coherence nitride-based superconducting qubits.
\end{abstract}


\end{frontmatter}

\section{Introduction}
In recent years, superconducting qubits have emerged as a promising platform for realizing fault-tolerant quantum computing owing to their design flexibility and high integration density. 
Because of tremendous advances in circuit architecture \cite{Koch2007-ec, Manucharyan2009-la} and materials \cite{Place2021-uy, Chang2013-aq}, the coherence time of state-of-the-art superconducting qubits has reached the millisecond regime \cite{Somoroff2023-at, Bland2025-bu} and gate fidelities have exceeded the break-even point \cite{Ni2023-ox, Google-Quantum-AI-and-Collaborators2025-er}. 
Superconducting qubits consist of Josephson junctions, planar capacitors, and readout resonators.
For planar capacitors and readout resonators, various oxidation-resistant superconductors, such as TiN \cite{Chang2013-aq}, Ta \cite{Place2021-uy, Bland2025-bu}, and capped Nb \cite{Bal2024-sc}, have been employed to suppress microwave dielectric losses originating from defects in native oxides formed on film surfaces \cite{Martinis2005-em, McDermott2009-zh, Muller2019-es}.
In contrast, because of their well-established fabrication process \cite{Dolan1977-zo, Costache2012-py}, Al/$\rm AlO_x$/Al Josephson junctions have been widely adopted in superconducting qubits.
Extensive process development based on Al/$\rm AlO_x$/Al Josephson junctions has been carried out toward large-scale integration, leading to highly reproducible fabrication processes \cite{Kreikebaum2020-ap}, post-fabrication critical current tuning techniques for correcting qubit frequencies \cite{Hertzberg2021-nh, Pappas2024-ph}, and lift-off-free fabrication processes compatible with mature CMOS technology \cite{Verjauw2022-yy, Van-Damme2024-gu}.

However, it has been pointed out that the coherence properties of qubits based on Al/$\rm AlO_x$/Al junctions are degraded by microscopic two-level systems (TLSs) in the amorphous aluminum oxide ($\rm AlO_x$) tunnel barrier, which originate from dangling bonds, crystal defects, and other structural disorders \cite{Martinis2005-em, Muller2019-es}.
To reduce the density of TLSs in Josephson junctions, one promising approach is the use of epitaxially grown Josephson junctions.
A reduction in TLS density was reported for phase qubits incorporating epitaxial Re/crystalline $\rm Al_2O_3$/polycrystalline Al junctions fabricated on C-plane sapphire substrates \cite{Oh2005-ps, Oh2006-mg}.
Fully epitaxial NbN/AlN/NbN Josephson junctions \cite{Wang2013-qp} are also promising alternatives to conventional Al/$\rm AlO_x$/Al Josephson junctions for reducing TLS density and enhancing the coherence properties of superconducting qubits.
An early study demonstrated phase qubits incorporating epitaxial NbN/AlN/NbN junctions with coherence times on the order of several microseconds \cite{Yu2002-cg}, comparable to the relatively long coherence times reported for phase qubits at that time \cite{Martinis2002-ob}. Subsequently, frequency-tunable transmon qubits based on these junctions were demonstrated on single-crystal MgO substrates, exhibiting an energy-relaxation time $T_1$ of approximately 500 ns \cite{Nakamura2011-ae}.
However, these coherence times were shorter than those of qubits based on Al/$\rm AlO_x$/Al junctions, primarily because they were limited by dielectric loss in the MgO substrate \cite{Qiu2014-wc, Watanabe2024-mq}.
To suppress dielectric loss from the MgO substrate, C-shunt flux qubits incorporating NbN/AlN/NbN junctions fabricated on high-resistivity Si substrates were demonstrated \cite{Sun2015-jt, Qiu2020-mq}. 
In these qubits, a TiN buffer layer was introduced to enable the epitaxial growth of NbN/AlN/NbN junctions on Si substrates despite the large lattice mismatch between Si and NbN \cite{Sun2015-jt}. 
These qubits achieved coherence times of several tens of microseconds \cite{Kim2021-ch}.

However, these trilayer junctions generally require interlayer insulation by depositing amorphous dielectric materials \cite{Qiu2020-mq, Kim2021-ch}.
In \cite{Kim2021-ch}, the fabrication process includes CVD deposition of $\rm SiO_2$ followed by chemical mechanical polishing (CMP) planarization.
These additional processing steps increase fabrication complexity while potentially reducing fabrication yield and increasing device-to-device variability, which may pose challenges for large-scale integration.
To avoid these issues, $\rm SiO_2$-free fabrication processes have been explored, with qubit lifetimes of tens of microseconds reported for native-oxide-passivated Nb/Al/$\rm AlO_x$/Nb trilayer junctions \cite{Sethi2025-gs} and several microseconds for flip-chip-bonded NbN/AlN/NbN junctions \cite{Wang2026-ct}.
However, eliminating $\rm SiO_2$ often comes at the cost of increased fabrication complexity, requiring additional processing steps or specialized device architectures.

Early qubits based on Nb/Al/$\rm AlO_x$/Nb junctions exhibited short coherence times \cite{Martinis2002-ob, Yu2005-hr, Weides2011-zp} due to dielectric losses arising from amorphous dielectric materials used for interlayer insulation \cite{Anferov2024-eg}.
Recently, the coherence times of qubits based on Nb junctions have been improved by more than two orders of magnitude through the development of a novel fabrication process \cite{Anferov2024-eg}.
In this process, a sacrificial $\rm SiO_2$ layer is anisotropically etched to form a self-aligned sidewall spacer structure \cite{Grimm2017-uq, Gronberg2017-gj}, leaving the spacer only where passivation is required. 
Compared with conventional fabrication processes, this approach significantly reduces both the volume of lossy amorphous $\rm SiO_2$ and fabrication complexity.
Furthermore, the sacrificial $\rm SiO_2$ is subsequently removed at the end of the fabrication process, thereby further reducing dielectric loss.

In this work, we developed sidewall spacer passivated epitaxial NbN/AlN/NbN Josephson junctions for superconducting qubit applications. As a result, excellent tunneling characteristics were obtained over a wide range of critical current densities ($J_{\rm c}$), from sub-$\rm A/cm^2$ to tens of $\rm kA/cm^2$.
We also confirmed that wet etching of the sacrificial $\rm SiO_2$ spacer did not degrade the junction characteristics.
These results establish a scalable fabrication process for epitaxial NbN/AlN/NbN junctions and demonstrate their potential as an alternative platform for high-coherence superconducting qubits.

\section{Device fabrication}

\begin{figure*}[t]
    \centering
    \includegraphics[width = \linewidth]{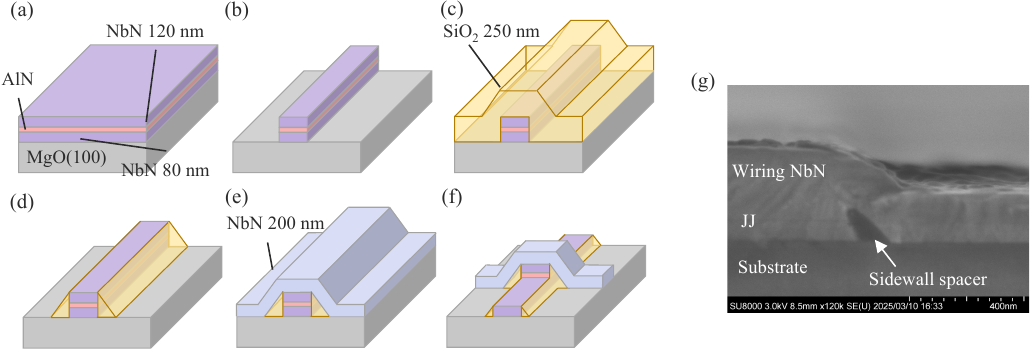}
    \caption{Fabrication process and cross-sectional SEM image of an NbN/AlN/NbN trilayer junction with the sidewall spacer structure: (a) An NbN/AlN/NbN trilayer is deposited on an MgO (100) substrate. (b) The trilayer is patterned and etched by chlorine-based ICP-RIE. (c) A $\rm SiO_2$ layer is deposited by CVD. (d) A sidewall spacer is formed by anisotropic etching of the $\rm SiO_2$ layer with a gas mixture of Ar, $\rm CF_4$, and $\rm CHF_3$. (e) An NbN wiring layer is deposited. (f) The wiring and counter-electrode layers are etched by fluorine-based RIE, thereby defining the junction area. (g) Cross-sectional SEM image of the fabricated junction.}
    \label{fig1}
\end{figure*}

The fabrication process for the NbN/AlN/NbN junctions is illustrated in Figure ~\ref{fig1}.
The NbN/AlN/NbN trilayer was deposited on a single-crystal MgO (100) substrate by reactive DC magnetron sputtering (Fig.~\ref{fig1} (a)).
MgO substrates were chosen to enable the direct epitaxial growth of the NbN/AlN/NbN trilayer.
Before trilayer deposition, the substrates were ultrasonically cleaned in organic solvents and ion-beam etched in the sputtering system to remove surface contaminants and promote adhesion and epitaxial growth.
The trilayer was deposited at ambient temperature without breaking the vacuum.
The critical current density $J_{\rm c}$, which is a key parameter determining the operating frequency of superconducting qubits, was controlled by varying the thickness of the AlN tunnel barrier.
To achieve precise control of the AlN thickness, the AlN barrier was sputtered in pure nitrogen gas at a low power density of approximately 0.3 $\rm W/cm^2$.
The deposition rates were approximately 1.1 $\rm nm/s$ for NbN and 0.05 $\rm nm/s$ for AlN.

The deposited trilayer was patterned using a maskless aligner and etched by inductively coupled plasma reactive ion etching (ICP-RIE) with a gas mixture of Ar, $\rm Cl_2$, and $\rm BCl_3$ (Fig.~\ref{fig1} (b)).
The chlorine-based gas mixture was selected to enable sequential etching of the NbN and AlN layers in a single ICP-RIE process.
To make electrical contact between the wiring and counter-electrode layers while preventing electrical shorts between the wiring and base-electrode layers, an insulating sidewall spacer was introduced.
A $\rm SiO_2$ layer with a thickness comparable to that of the patterned trilayer was deposited by plasma-enhanced chemical vapor deposition (PECVD) at a substrate temperature of 280${}^\circ$C (Fig.~\ref{fig1} (c)).
The $\rm SiO_2$ layer was anisotropically etched by ICP-RIE with a gas mixture of Ar, $\rm CF_4$, and $\rm CHF_3$ to form sidewall spacers along the junction sidewalls (Fig.~\ref{fig1} (d)).
After $\rm SiO_2$ etching, residues and surface contaminants were removed from the counter-electrode layer by ion-beam etching, and the NbN wiring layer was deposited by sputtering (Fig.~\ref{fig1} (e)).

The deposited wiring layer was patterned and etched by reactive ion etching (RIE) with $\rm CF_4$ gas (Fig.~\ref{fig1} (f)). 
In this step, the counter-electrode layer outside the junction was etched simultaneously, thereby defining the junction area.
A cross-sectional scanning electron microscopy (SEM) image of the fabricated junction is shown in Fig.~\ref{fig1} (g).
Because TLSs in amorphous $\rm SiO_2$ can contribute to dielectric loss at low microwave powers and temperatures, wet etching of the $\rm SiO_2$ sidewall spacer was considered as a means of reducing the amount of amorphous dielectric material. For selected samples, current-voltage characteristics were measured before and after buffered hydrofluoric acid (BHF) treatment to evaluate the effect of the treatment on the junction characteristics.

\section{Result and discussion}

\begin{figure*}[t]
    \centering
    \includegraphics[width = 0.9\linewidth]{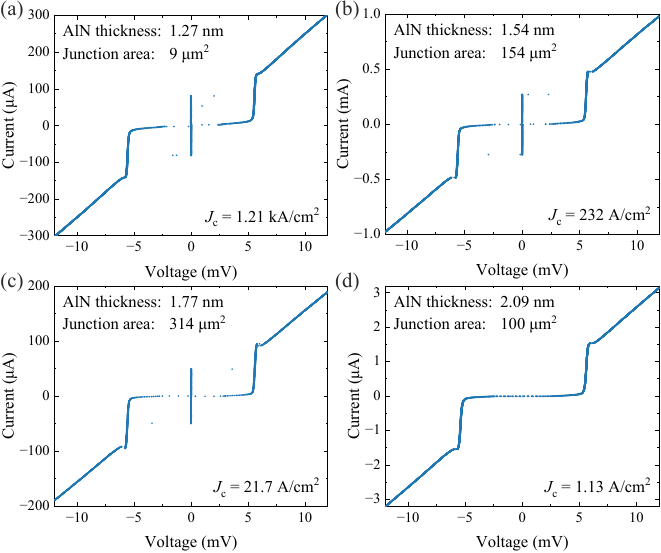}
    \caption{Current–voltage characteristics of NbN/AlN/NbN tunnel junctions with different AlN barrier thicknesses, measured at 2.3 K: (a) 1.27 nm, (b) 1.54 nm, (c) 1.77 nm, and (d) 2.09 nm. AlN thicknesses were estimated from the AlN deposition rate and deposition time.}
    \label{fig2}
\end{figure*}

\begin{table*}[t]
\caption{Electrical parameters extracted from the current–voltage characteristics shown in Figure~\ref{fig2}.}
\label{tab1}
\centering
\begin{tabular}{lcccccccc}
\hline
Junction &
\begin{tabular}[c]{@{}c@{}}$I_{\rm c}$\\ ($\rm \mu A$)\end{tabular} &
\begin{tabular}[c]{@{}c@{}}$J_{\rm c}^{\rm th}$\\ ($\rm A/cm^2$)\end{tabular} &
\begin{tabular}[c]{@{}c@{}}$V_{\rm g}$\\ ($\rm mV$)\end{tabular} &
\begin{tabular}[c]{@{}c@{}}$R_{\rm N}$\\ ($\rm \Omega$)\end{tabular} &
\begin{tabular}[c]{@{}c@{}}$R_{\rm sg}$\\ ($\rm \Omega$)\end{tabular} &
$R_{\rm sg}/R_{\rm N}$ &
\begin{tabular}[c]{@{}c@{}}$I_{\rm c}R_{\rm N}$\\ ($\rm mV$)\end{tabular} &
\begin{tabular}[c]{@{}c@{}}$I_{\rm c}R_{\rm N}\ (\rm theory)$\\ ($\rm mV$)\end{tabular} \\
\hline
a & 81.4 & 1.21k & 5.56 & 40.0& 673 & 16.8 & 3.25 & 4.37\\
b & 273 & 232 & 5.58 & 12.3 & 460 & 37.4 & 3.35 & 4.38\\
c & 64.3 & 21.7 & 5.57 & 64.3 & > 5.7k & > 89 & 3.18 & 4.37\\
d & -- & 1.13 & 5.53 & 3.83k & 385k & 101 & -- & 4.34\\
\hline
\end{tabular}
\end{table*}

The fabricated samples were cooled to 2.3 K in a Gifford–McMahon (GM) cryocooler, and their current–voltage characteristics were measured.
Figure~\ref{fig2} shows the measured current–voltage characteristics of the fabricated junctions with different AlN barrier thicknesses, $d_{\rm AlN}$.
Table~\ref{tab1} summarizes the junction parameters.
The critical current $I_{\rm c}$ listed in the second column of Table~\ref{tab1} was directly obtained from the measured current–voltage characteristics.
In contrast, the theoretical critical current density $J_{\rm c}^{\rm th}$, listed in the third column, was calculated using the Ambegaokar–Baratoff relation \cite{Ambegaokar1963-la} as $J_{\rm c}^{\rm th}=\pi V_{\rm g}/(4R_{\rm N} A)$, where $V_{\rm g}$ is the gap voltage determined from the voltage at which the differential conductance (d$I$/d$V$) reaches its maximum, $R_{\rm N}$ is the normal resistance measured at 10 mV, and $A$ is the nominal junction area.
This method was employed because direct measurement of true $I_{\rm c}$ is difficult owing to environmental noise and magnetic flux trapping.
The junction quality factor was evaluated as $R_{\rm sg}/R_{\rm N}$, where $R_{\rm sg}$ is the subgap resistance measured at 3 mV.
The theoretical $I_{\rm c}R_{\rm N}$ product ($I_{\rm c}R_{\rm N}$ (theory)) was calculated as $I_{\rm c}R_{\rm N}=\pi V_{\rm g}/4$. 

All fabricated NbN/AlN/NbN junctions with different $J_{\rm{c}}$ exhibited excellent electrical characteristics, characterized by a large gap voltage of approximately 5.6 mV and small subgap leakage currents.
In addition, the junction quality factor $R_{\rm sg}/R_{\rm N}$ was approximately 100 for the junctions with $J_{\rm{c}}$ of 21.7 and 1.13 $\rm A/cm^2$, and exceeded 10 for the junctions with $J_{\rm{c}}$ of 232 and 1.21k $\rm A/cm^2$.
The obtained gap voltages and subgap leakage currents were comparable to those previously reported for NbN/AlN/NbN junctions fabricated using different processes \cite{Wang2013-qp}.
These results indicate that the fabrication process applied in this work does not degrade the electrical characteristics of the junctions.

\begin{figure}[t]
    \centering
    \includegraphics[width = 0.9\linewidth]{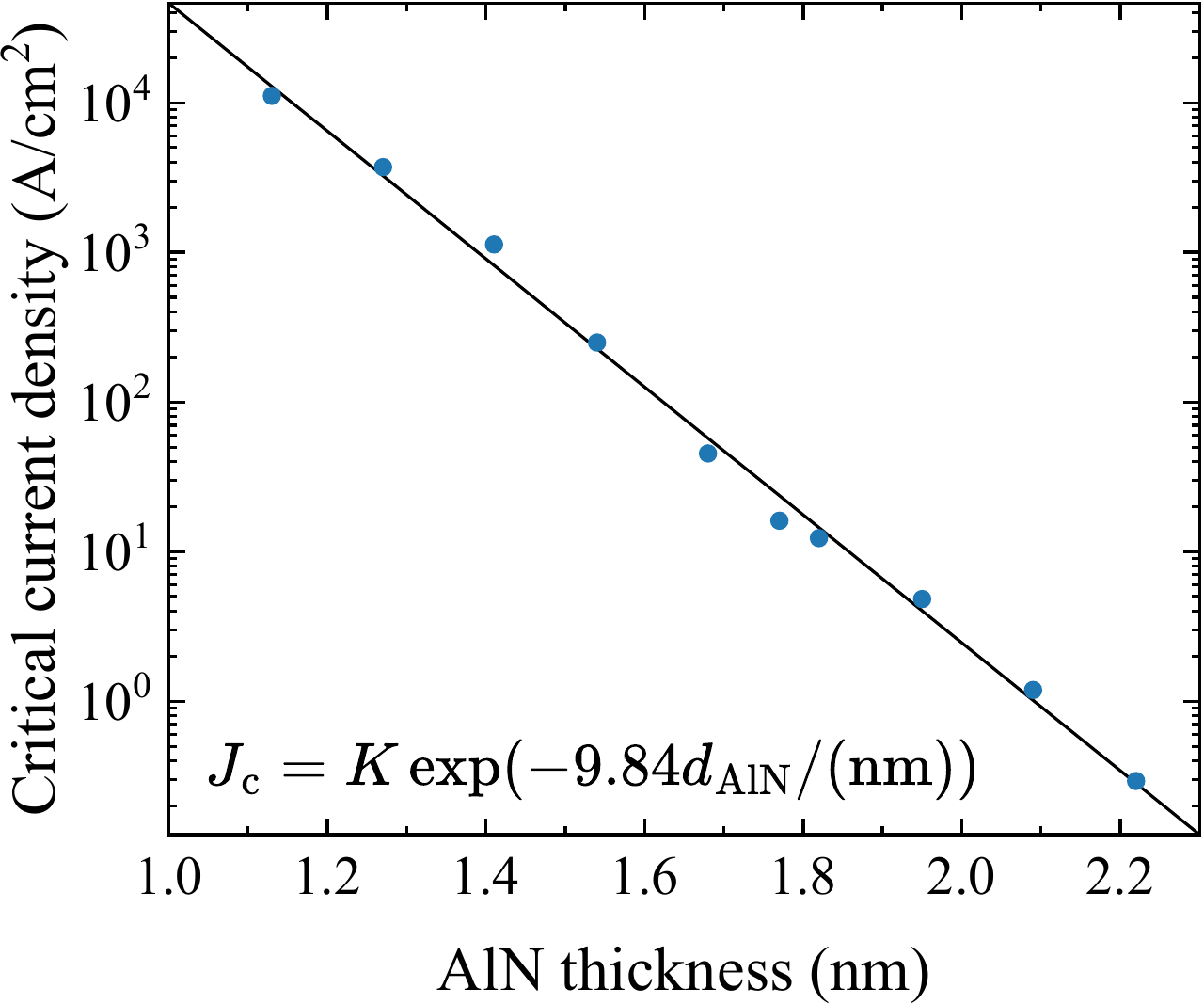}
    \caption{Dependence of the critical current density $J_{\rm c}$ on the AlN barrier thickness $d_{\rm AlN}$. The values of $d_{\rm AlN}$ were estimated from the AlN deposition rate and deposition time. The solid line represents an exponential fit to the experimental data.}
    \label{fig3}
\end{figure}

Precise control of $J_{\rm c}$ is essential for superconducting qubit applications.
For transmon qubits \cite{Koch2007-ec}, the qubit transition frequency $f_{01}$ is approximately given by $f_{01}=(\sqrt{8E_{\rm J}E_{\rm C}}-E_{\rm C})/h$, where the Josephson energy $E_{\rm J}$ and charging energy $E_{\rm C}$ are defined as $E_{\rm J}=\Phi_0 I_{\rm c}/2\pi$ and $E_{\rm C}=e^2/2C$, respectively, and $C$ is the total qubit capacitance.
Because $E_{\rm J}$ is determined by $I_{\rm c}=J_{\rm c}A$, precise control of both $J_{\rm c}$ and the junction area is essential for accurately setting the qubit transition frequency.
In NbN/AlN/NbN junctions, $J_{\rm c}$ is controlled by the thickness of the AlN tunnel barrier. 
Therefore, the AlN barrier was deposited at a low rate of 0.05 $\rm nm/s$ to enable precise control of its thickness.
Figure~\ref{fig3} shows the dependence of $J_{\rm c}$ on the AlN barrier thickness $d_{\rm AlN}$.
The AlN barrier thickness $d_{\rm AlN}$ was estimated from the AlN deposition rate and deposition time.
As shown in Fig.~\ref{fig3}, $J_{\rm c}$ increased exponentially with decreasing $d_{\rm AlN}$ and was fitted by $J_{\rm c}=K\exp(-9.84d_{\rm AlN}/(\rm nm))$, where $K$ is a fitting constant.
The exponential dependence of $J_{\rm c}$ on the AlN barrier thickness arises from the exponential dependence of the tunneling probability on the barrier thickness.
The barrier height estimated using the Simmons model \cite{Simmons1963-qo} was approximately 1.0 eV, comparable to previously reported values for NbN/AlN/NbN junctions fabricated on MgO substrates \cite{Wang2013-qp}.
In this work, $J_{\rm c}$ was controlled over a wide range from 0.3 to $10^4$ $\rm A/cm^2$ by precisely adjusting the AlN barrier thickness.

The required $J_{\rm{c}}$ depends on the type of transmon qubit. 
Conventional transmon qubits are typically designed with $J_{\rm c}$ of several tens of $\rm A/cm^2$ \cite{Koch2007-ec}, whereas transmons operating at higher frequencies require higher $J_{\rm c}$ exceeding $10^3$ $\rm A/cm^2$ \cite{Anferov2025-an} to suppress thermal excitation at elevated temperatures.
In contrast, merged-element transmons with a reduced footprint are designed with a lower $J_{\rm c}$ of approximately 1 $\rm A/cm^2$ \cite{Zhao2020-ym, Mamin2021-iw}.
The wide tunability of $J_{\rm c}$ demonstrated in this work, together with the low subgap leakage of the fabricated junctions, indicates the potential applicability of these junctions to various types of superconducting qubits.

\begin{figure}[t]
    \centering
    \includegraphics[width = 0.9\linewidth]{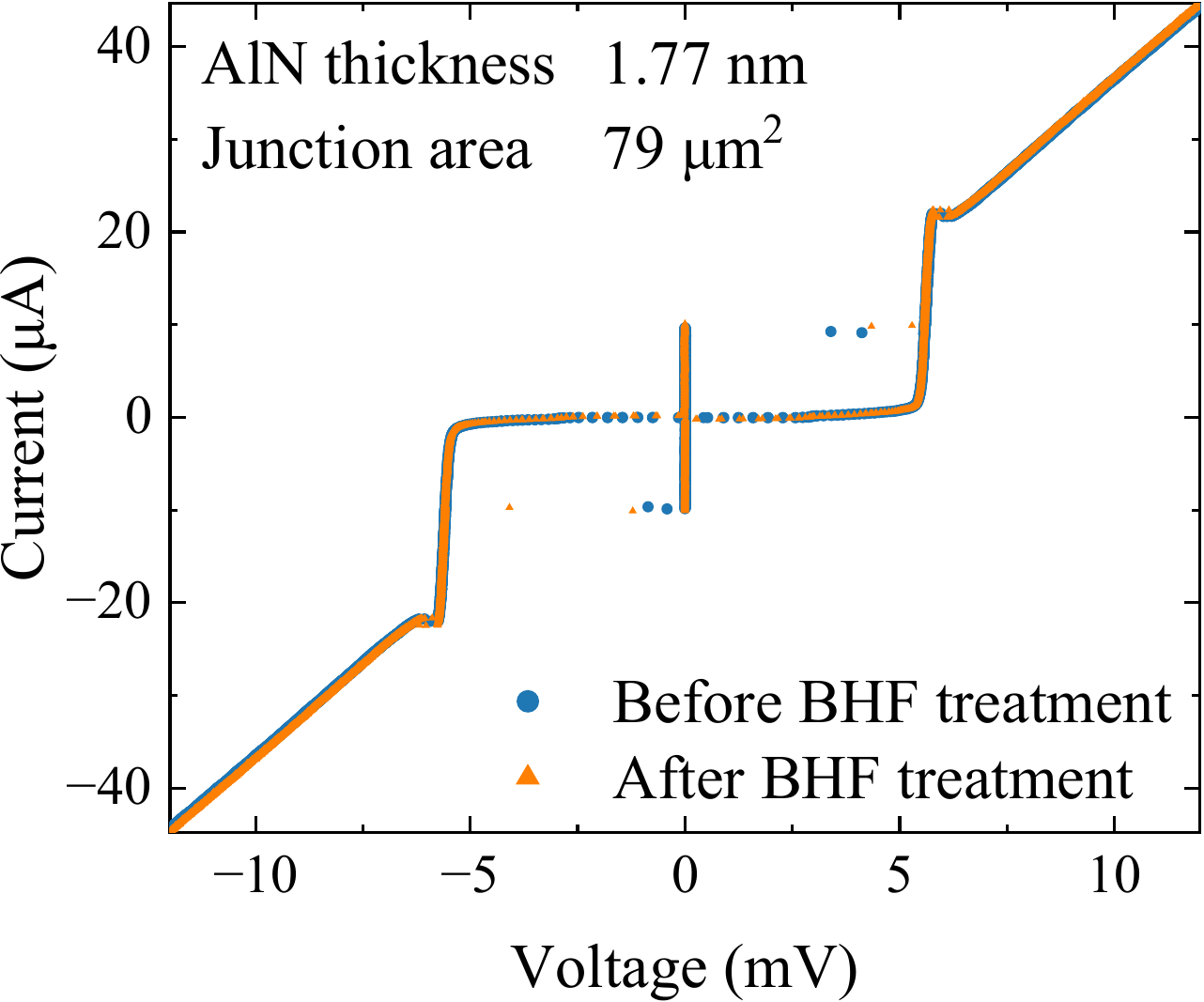}
    \caption{Current–voltage characteristics of an NbN/AlN/NbN junction measured at 2.3 K before and after BHF treatment. The blue and orange symbols represent the data measured before and after BHF treatment, respectively.}
    \label{fig4}
\end{figure}

Unlike conventional Al/AlO$_{\rm x}$/Al junctions fabricated using the Dolan-bridge lift-off process \cite{Dolan1977-zo}, fabrication processes for trilayer junctions generally require a sacrificial $\rm SiO_2$ layer to electrically isolate the base-electrode layer from the wiring layer, as in the process used in this work.
Amorphous dielectrics are known to degrade the coherence of superconducting qubits through dielectric loss associated with TLSs \cite{Martinis2005-em}.
Therefore, buffered hydrofluoric acid (BHF) is commonly used at the end of the fabrication process to etch the sacrificial $\rm SiO_2$ layer and native surface oxides \cite{Kim2021-ch, Anferov2024-eg}.
However, for Nb/Al/$\rm AlO_x$/Nb trilayer junctions, even brief etching in a solution of ammonium fluoride and acetic acid has been reported to partially dissolve the Al/$\rm AlO_x$ tunnel barrier from the junction edge, thereby altering the junction characteristics \cite{Anferov2025-an}.
Therefore, we investigated the effect of wet etching of the $\rm SiO_2$ layer on the electrical characteristics of NbN/AlN/NbN junctions.
Figure~\ref{fig4} shows the current–voltage characteristics measured before and after BHF treatment.
The blue and orange symbols represent the current–voltage characteristics measured before and after BHF treatment, respectively.
No significant differences were observed between the current–voltage characteristics measured before and after BHF treatment, and the large gap voltage and low subgap leakage current were maintained after treatment.
These results indicate that BHF treatment does not degrade the electrical characteristics of NbN/AlN/NbN junctions.

\section{Conclusion}
In this work, we applied a sidewall-spacer fabrication process to epitaxial NbN/AlN/NbN Josephson junctions and evaluated their electrical characteristics for superconducting qubit applications. 
The fabricated junctions exhibited a large gap voltage of approximately 5.6 mV and low subgap leakage currents. 
The junction quality factor $R_{\rm sg}/R_{\rm N}$ was approximately 100 for the junctions with $J_{\rm c}$ values of 1.13 and 21.7 $\rm A/cm^2$, and exceeded 10 for the junction with $J_{\rm c}$ of 232 and 1.21k $\rm A/cm^2$. 
By precisely adjusting the AlN barrier thickness, high-quality junctions were fabricated over a wide $J_{\rm c}$ range from 0.3 to $10^4$ $\rm A/cm^2$, covering the values required for various types of transmon qubits.
In addition, measurements performed before and after BHF treatment indicated that the wet-etching process did not degrade the electrical characteristics of the junctions.
These results provide a versatile junction platform for superconducting qubits employing nitride superconductors and support the development of devices with a broad range of operating frequencies and circuit architectures.

\section*{CRediT authorship contribution statement}
\textbf{Koki Honda:} Writing – review\&editing, Writing – original draft, Visualization, Validation, Resources, Methodology, Investigation, Funding Acquisition, Data Curation, Conceptualization. \textbf{Daiki Kurihara:}Writing – review\&editing, Software, Resources, Data Curation. \textbf{Hiroki Kutsuma:} Writing – review\&editing, Resources, Supervision, Funding Acquisition. \textbf{Hirotaka Terai:} Writing – review\&editing, Supervision, Resources. \textbf{Taro Yamashita:} Writing – review\&editing, Visualization, Supervision, Resources, Project Administration, Methodology, Funding Acquisition, Conceptualization.

\section*{Declaration of competing interest}
The authors declare that they have no known competing financial interests or personal relationships that could have appeared to influence the work reported in this paper.

\section*{Acknowledgments}
The authors would like to thank K. Takaki and M. Setoura for their assistance with sample fabrication.
This work was partly supported by Grant-in-Aid for JSPS Fellows (26KJ0584), JST CREST (JPMJCR24I5), Grant-in-Aid for Scientific Research (S) (25H00406), JST Moonshot R\&D Program (JPMJMS2067), and Advanced Research Infrastructure for Materials and Nanotechnology in Japan (ARIM) of MEXT (JPMXP1226TU0021). T.Y. and H.K. acknowledge Center for Heterogeneous Quantum/Material Fusion Technologies, Center for Key Interdisciplinary Research, Tohoku University.

 \bibliographystyle{elsarticle-num} 
 \bibliography{main}

\end{document}